\documentstyle[12pt]{article}

\begin{document}

\renewcommand{\thesection}{\arabic{section}.}
\renewcommand{\theequation}{\thesection \arabic{equation}}
\newcommand{\scs}{\setcounter{equation}{0} \setcounter{section}}
\def\req#1{(\ref{#1})}
\newcommand{\be}{\begin{equation}} \newcommand{\ee}{\end{equation}}
\newcommand{\ba}{\begin{eqnarray}} \newcommand{\ea}{\end{eqnarray}}
\newcommand{\la}{\label} \newcommand{\nb}{\normalsize\bf}
\newcommand{\lb}{\large\bf} \newcommand{\vol}{\hbox{Vol}}
\newcommand{\bb} {\bibitem} \newcommand{\np} {{\it Nucl. Phys. }}
\newcommand{\pl} {{\it Phys. Lett. }}
\newcommand{\pr} {{\it Phys. Rev. }} \newcommand{\mpl} {{\it Mod. Phys. Lett.
}}
\newcommand{\sg}{{\sqrt g}} \newcommand{\sqhat}{{\sqrt{\hat g}}}
\newcommand{\sqphi}{{\sqrt{\hat g}} e^\phi} \newcommand{\sqalpha}{{\sqrt{\hat
g}}e^{\alpha\phi}}
\newcommand{\tp}{\cos px\ e^{(p-{\sqrt2})\phi}} \newcommand{\stwo}{{\sqrt2}}

\begin{titlepage}
\renewcommand{\thefootnote}{\fnsymbol{footnote}}

\hfill PUPT--1542

\hfill hep-th/9506118

\vspace{.4truein}
\begin{center}
 {\LARGE RG flow on random surfaces with handles}\vskip5mm
 {\LARGE and closed string field theory}
 \end{center}
\vspace{.7truein}

 \begin{center}

 Christof Schmidhuber\footnote{schmidhu@puhep1.princeton.edu

 \ \ on leave of absence from the Institute for Theoretical Physics, University
of Bern, Switzerland.}

 \vskip5mm

 {\it Joseph Henry Laboratories}

 {\it Princeton University}

 {\it Princeton, NJ 08544, USA}

 \end{center}

\vspace{.7truein}
\begin{abstract}

The renormalization group flow in two-dimensional field theories that are
coupled to gravity has unusual features:
First, the flow equations are second order in derivatives.
Second, in the presence of handles the flow has quantum mechanical properties.
Third, the beta functions contain the elementary higher-genus vertices of
closed string
field theory. This is demonstrated at simple examples and is applied to derive
various
results about gravitationally dressed beta functions. The possibility of
interpreting
closed string field theory as the theory of the renormalization
group on random surfaces with random topology
is considered.

\end{abstract}
 \renewcommand{\thefootnote}{\arabic{footnote}}
 \setcounter{footnote}{0}
\end{titlepage}

{}\subsection*{1. Introduction}\scs{1}

Recently there have been several studies of the flow of coupling constants in
two-dimensional field
theories that are coupled to gravity \cite{pol,kpz,ddk}. Among the results are
the phase diagram
of the sine-Gordon model coupled to gravity \cite{sch} and the gravitational
dressing of
beta function coefficients for bosonic \cite{sch,pkk,amb,tani,dorn} and
supersymmetric theories \cite{gri}.
These results can be obtained either in light-cone gauge or in conformal gauge.
While the light-cone gauge approach is more rigorous and uses standard
techniques of field theory,
the conformal gauge approach seems to always yield the same results with less
effort.
In this paper it is shown that it can also be easily extended to
higher-genus surfaces.

We consider two-dimensional conformal field theories perturbed by scaling
operators $\Phi_i$
with coupling constants $\lambda^i$. The action is
\ba S=S_{cft}+\lambda^i\int d^2\xi\ \sg\ \Phi_i.\la{jane}\ea
Without gravity, the flow of coupling constants is then described by the flow
equations
\ba\dot\lambda^i(\tau)\ \ =\ \ \beta^i(\lambda^j)\la{laura}\ea
with beta functions $\beta^i$ and ``renormalization group time'' $\tau$.
When gravity is ``turned on'' and the world-sheet topology is allowed to
fluctuate, this flow is modified.
Three types of modifications will be discussed here:
\begin{enumerate}
\item
The time-derivative in (\ref{laura}) is replaced by a simple second-order
derivative
operator.\footnote{This has been pointed out previously
e.g. in \cite{das}.}
\item
Classical flow trajectories $\lambda^i(\tau)$ are replaced by ``wave packets''
that can spread in the space of coupling constants. Equivalently, the
$\lambda^i$ are replaced
by quantum operators.
\item
The beta functions on the right-hand side of (\ref{laura}) are modified by the
elementary higher-genus vertices of closed string field theory.
\end{enumerate}
These modifications will be derived and discussed
at the examples of the minimal models and the $c=1$ model on a circle up to
cubic order in $\lambda,\kappa$,
where $\kappa^2$ is the topological coupling constant.
Our motivation for studying the renormalization group flow on random surfaces
is two-fold.
First, we hope that the results about phase diagrams and gravitationally
dressed coefficients mentioned above will be followed by similar
results about the critical behavior of other, more physical systems of random
surfaces,
like perhaps the $3d$ Ising model or QCD.

Second, points 2 and 3 extend to higher genus the observation
\cite{das,polch,tseyt,suss,poLH} that
renormalization group trajectories
in $2d$ field theories coupled to gravity can be regarded as classical
solutions of string theory, where the $\lambda^i$ represent
the target space fields, $\tau$ is related to time and the flow equations are
related to the string
equations of motion (hence point 1).
Reversing the viewpoint, one is tempted to {\it define} perturbative string
theory in terms of
the flow on (super) random surfaces. Our hope is that this definition, which is
made more explicit
towards the end of this paper, can be naturally extended to a nonperturbative
one.
\vskip2cm

The line of argument in this paper is as follows.
Coupling constants flow because of logarithmic divergences that make the
renormalized coupling constants sensitive to the cutoff scale. As is
well-known, such
divergences come from the boundary of moduli space, i.e., from pinched surfaces
whose pinching radius
$r$ is restricted by the short--distance cutoff $a$ on the world--sheet: $r\ge
a$.
Three types of pinched surfaces (nodes) can be distinguished:
\vskip3mm

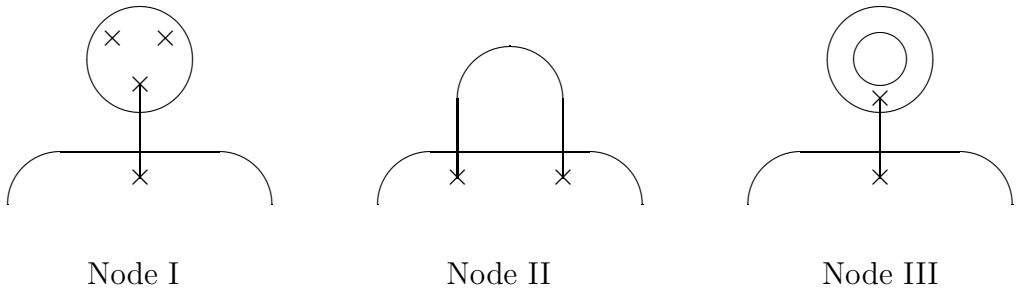
\begin{picture}(400,120)
\put(20,20){ \begin{picture}(100,100) \put(50,20){\oval(100,40)[t]}
\put(50,75){\circle{40}} \put(50,30){\line(0,1){35}}
  \put(45.4,27.5){$\times$}  \put(45.4,62.5){$\times$}  \put(35,80){$\times$}
\put(55,80){$\times$}  \put(30,-10){Node I}  \end{picture}}

\put(300,20){ \begin{picture}(100,100) \put(50,20){\oval(100,40)[t]}
\put(50,75){\circle{40}} \put(50,75){\circle{18}}
  \put(50,30){\line(0,1){30}} \put(45.4,27.5){$\times$}
\put(45.4,57.5){$\times$}  \put(28,-10){Node III}  \end{picture} }

\put(160,20){ \begin{picture}(100,100) \put(50,20){\oval(100,40)[t]}
\put(50,30){\oval(40,100)[t]}
  \put(25.4,27.5){$\times$}  \put(65.4,27.5){$\times$}  \put(26,-10){Node II}
\end{picture} }

\end{picture}

Each ``$\times$'' represents an operator insertion and each line a pinch (the
`ring' in the last diagram
represents a torus).
Node I is already present on genus zero surfaces. In the absence of gravity, it
leads to the standard
quadratic beta function coefficients. In section 2 it is first reviewed how
gravity modifies these coefficients in the conformal gauge approach. The
relation with the string equations
of motion is then explained following \cite{st} and is used to derive the
modification of cubic beta function coefficients by gravity.

Node II represents a pinched handle. As is demonstrated in section 3, all scale
dependence coming from
massive modes (as defined later) propagating through node II can be absorbed in
a running topological
coupling constant. But massless modes lead to a bilocal scale dependence that
can only be absorbed by
``quantizing'' the flow (the precise meaning of this will become clear).
Some observations about various issues including background (in) dependence are
also  made in section 3.

Node III is responsible for the well-known Fischler-Susskind mechanism
\cite{fs}.
In section 4 its effect on the flow of the radius in the $c=1$ model on a
circle is discussed.
It is pointed out that the corresponding poles precisely agree with those seen
in matrix model amplitudes.
Now, the scale dependence of part of node III is already absorbed by
quantizing the flow to account for node II, as seen in section 5.
Only the remainder, an elementary genus-one closed string field vertex, then
contributes to the beta
functions.

Section 6 contains some suggestions about the flow in supersymmetric theories
with central charge $\hat c\ge9$ coupled to supergravity, about the possibility
of
tunneling of the flow and about viewing closed string field theory as the
theory of the flow
in unitary theories on random surfaces. Section 6 also contains a summary of
our results.

For simplicity, in this paper attention is restricted to states with equal
left-
and right-moving momenta (thus excluding winding modes) that also do not
contain ghost operators.
Also, we will not worry about possible contributions of gravitational
descendents from boundaries of moduli space, since it is expected (from
BRST invariance in string theory) that the sum of these contributions cancels.
It is understood that a rigorous discussion should include an analysis of the
full BRST cohomology
and its interactions in the spirit of \cite{polski,wizwi}.
However, we do not expect that this will modify our basic conclusions.

{}\subsection*{2. Node I and and second-order flow equations}\scs{2}

In this section it is reviewed how the standard renormalization group flow
is modified by gravity on a genus zero surface. The ``gravitational dressing''
of cubic beta function coefficients is also derived here, in qualitative
agreement with the
independent result \cite{dorn}.

 {}\subsubsection*{2.1. Renormalization group flow in conformal gauge}

Since conformal gauge will be used to study the flow, let us
first recall at the example of node I how the dressing of beta functions is
derived in this gauge.
It is done in two steps. First, the effective action for the conformal factor
is constructed.
Then constant shifts of the conformal factor are absorbed in running coupling
constants.
It will be assumed that the cosmological constant, which is small in the
ultraviolet, does not affect
the short-distance effects that are responsible for the flow of coupling
constants. This
leads to agreement with matrix model results (see below).

Theory (\ref{jane}) coupled to gravity can be described in conformal gauge by a
conformal field theory with an additional field, the conformal factor $\phi$ of
the world--sheet metric.
The first step is to write down the most general local renormalizable action
for the combined theory,
order by order in the coupling constants $\lambda^i$ \cite{ddk} (setting
$\alpha'=2$):
\begin{eqnarray}
  S=S_{\hbox{cft}}+{1\over{8\pi}}\int  d^2\xi\ {\sqrt{\hat g}}\{\ (\partial
\phi)^2
       &-& Q R^{(2)} \phi\ +\ \hbox{ghosts}\ \}\la{anne}\\
       &+& \lambda^i \int\Phi_i e^{\alpha_i \phi}\la{anne1}\\
       &+& \lambda^j\lambda^k\int X_{jk}\la{anne2}\\
       &+& \lambda^j\lambda^k\lambda^l\int X_{jkl} + ...\ .\la{jeanne}
\end{eqnarray}
Here, $X_{jk}(\phi), X_{jkl}(\phi)$ are operators to be determined below.
$\hat g$ is a fictitious, arbitrarily chosen background metric that nothing
physical can depend on.
In particular, the combined theory, including all of its correlation functions
must be scale invariant. This is the guiding principle that determines the
coefficients $Q$, $\alpha_i$
and the operators $X_{jk}, X_{jkl}$ order by order in $\lambda$: to
zeroth order, scale invariance determines
$$Q={1\over3}{\sqrt{25-c}}$$
to make the total central charge zero. To linear order in $\lambda$, scale
invariance requires
\ba \alpha_i^\pm = -{Q\over2}\pm\omega_i\ \ \ \ \ \hbox{with}\ \ \ \ \
\omega_i={\sqrt{h_i-2+{Q^2\over4}}}
\ ,\la{rob}\ea
where $h_i$ is the scaling dimension of $\Phi_i$,
such that the operators
$$V_i\equiv\Phi_ie^{\alpha_i^+ \phi}\ \ ,\ \ \ \
\bar V_i\equiv\Phi_ie^{\alpha_i^- \phi}\ =\ \Phi_ie^{(-Q-\alpha_i^+) \phi}$$
have dimension two.
$\omega_i$ can be thought of as (imaginary) frequency. The operator $\bar V_i$
is usually
assumed to ``not exist'' \cite{seipol} and will be ignored here.\footnote{$\bar
V_i$ will play a role in
the presence of handles, as will be seen in later sections.}
If $\Phi_i$ is almost marginal, $\omega_i\sim{Q\over2}$ and $\alpha_i^+\sim0$.
To quadratic order in $\lambda$, scale invariance then determines $X_{jk}$,
which in this case is a universal operator. One finds \cite{sch}:
\ba X_{jk} \ =\ {\pi\over 2\omega_i}
\ c^i_{jk}\ \phi V_i\ .\la{ted}\ea
Indeed, this term is needed to ensure scale invariance of the two--point
function,
i.e., of the second derivative of the partition function with respect to the
coupling constants:
\ba{\delta^2 \over \delta\lambda^j\delta\lambda^k}\ Z\ \ \sim\ \
<e^{-S}\int V_j\int V_k>\ -\ {\pi\over{2\omega_i}}c^i_{jk}\ <e^{-S}\int \phi
V_i>\ .\la{sylvia}\ea
Namely, the short--distance singularity in the operator product expansion (OPE)
of $V_j$ with $V_k$ must be regularized, e.g. by setting a minimal
distance $\hat a$ between the operators. Then the first term in (\ref{sylvia})
contains a divergent part
\ba -\ \pi c^i_{jk} (\log \hat a)\ <e^{-I}\int V_i> \la{jimmy} \ea
that depends on the fictitious scale through the fictitious cutoff $\hat a$.
Since
\ba(L_0+\bar L_0-2)\ \phi V_i = -2\omega_i V_i,\la{dora}\ea
the second term in (\ref{sylvia}) exactly cancels this scale dependence
($L_0+\bar L_0-2$ is the generator of scale transformations).
This situation, where $V_j$ approaches $V_k$, is conformally equivalent to
node I, where a long cylinder of length $L\sim\log\hat a$ connects two separate
surfaces.
For $\alpha_i=0$, the interaction terms (\ref{anne1}+\ref{anne2}) become
\ba \lambda^i\Phi_i + {\pi\over{Q}}\ c^i_{jk}\lambda^j\lambda^k\ \Phi_i\ \phi.
\la{mary}\ea

The second step is to read off the renormalization group flow ``in the presence
of gravity'' from the action.
By construction, there is no flow with respect to the fictitious background
scale defined by $\sqhat$.
But since $\sg=\sqalpha$ defines the physical scale, a constant shift
\ba\phi\rightarrow \phi+{2\over\alpha}\tau \ \ \ ,\ \ \ \  \
\alpha=-{Q\over2}+{\sqrt{{Q^2\over4}-2}}\la{linda}\ea
corresponds to a scale transformation or -- more
precisely -- to a rescaling of the {\it physical} cutoff $a\ \rightarrow\
ae^\tau$. The crucial point here
is that $\phi$ lives on a half-line, bounded by the physical cutoff:
$\phi\le(\log a)/\alpha$ \cite{polch}.
Otherwise a constant shift of $\phi$ would be trivial since $\phi$ is
integrated over.
Such a shift of $\phi$ can now be absorbed in running coupling constants
$\lambda^i(\tau)$,
where $\tau\rightarrow\infty$ corresponds to the ultraviolet.
E.g., in (\ref{mary}) the shift (\ref{linda}) induces
$$ \ \tau\ {2\pi\over \alpha Q} c^i_{jk} \lambda^j\lambda^k\ \Phi_i.$$
This term can be absorbed in (\ref{mary}) to order $\lambda^2$ by defining
\cite{sch}
\ba\lambda^i(\tau) \sim \
 -\ \tau\ {2\pi\over Q\alpha}\  c^i_{jk}\lambda^j\lambda^k\ \ \ \rightarrow\ \
\
\dot\lambda^i\ \sim \
 - {2\over Q\alpha}\ \pi c^i_{jk}\lambda^j\lambda^k.\la{layla}\ea
By comparison, the flow without gravity is given by
\ba
\dot\lambda^i\ \sim \
 \pi c^i_{jk}\lambda^j\lambda^k\
 +\ \pi d^i_{jkl}\lambda^j\lambda^k\lambda^l\ +\ ...\la{lay}.\ea
The ``gravitational dressing factor'' $-2/(Q\alpha)$ in (\ref{layla})
of the universal quadratic term indeed agrees with the light cone gauge result
\cite{pkk}.
The method presented here can also be used to derive the phase diagram of the
sine-Gordon model coupled to
gravity \cite{sch}, in agreement with matrix model results \cite{gkl,moo}.

 {}\subsubsection*{2.2. Dressing of the $\lambda^3$ coefficients}

To determine the gravitational dressing of the cubic beta function coefficients
$d^i_{jkl}$, it is useful to note that
the scale invariance conditions for the effective action
(\ref{anne}--\ref{jeanne})
are just the string equations of motion \cite{frats,cfmp} with $\phi$ playing
the role of (euclidean) time.

Here it is assumed that the two-dimensional matter theory can be
formulated as a sigma model; the dressed ``matter'' operators $\Phi_i$ then
correspond to target space
gravitons or almost marginal tachyon perturbations.
E.g., the $q$-th minimal model with large $q$ (large $q$ ensures the existence
of almost marginal perturbations)
can be described by a scalar field $x$ with a Landau-Ginzburg potential
$T_q(x)$ \cite{zam}.
Perturbations around the fixed point can be expanded in a complete set of
scaling operators $T_{q,k}(x)$.
After coupling to gravity one obtains a conformal field theory with $x$ and
$\phi$ as target space coordinates,
a dilaton background $\Phi$ and a tachyon background $T$ that can be expanded
as
$$ T(x,\phi) = T_q(x) + \lambda^k(\phi) T_{q,k}(x).$$
The trajectory $\lambda^k(\phi)$ then describes a classical string solution
that asymptotically
approaches the static solution $\lambda^k(\phi)=0$.
Graviton perturbations $G_{\mu\nu}(x,\phi)$ can be treated similarly if one
picks the gauge $G_{\phi\phi}=1,G_{x\phi}=0$
as in (\ref{anne}-\ref{jeanne}). Expanding the string equations of motion in
$\lambda^i$, one finds the universal form (setting $\alpha'=2$) \cite{st}:
\ba O(\dot\lambda^2) + \ddot{\lambda}^i+{Q}\dot{\lambda}^i= \beta^i,\ \ \ \ \
Q^2 = {25-c \over3} + O(\lambda^2)\ .  \label{sue}\ea
Here, the dot means derivative with respect to $\phi$.
The first equation is the graviton or tachyon equation
while the second equation is the equation for the (shifted) dilaton zero mode,
i.e., the $x$-independent part $\varphi_0(\phi)$ of the (shifted) dilaton
$\varphi(x,\phi)\equiv2\Phi-{\sqrt G}$ ($Q$ is defined as
$-\dot\varphi_0(\phi)$; to order $\lambda^2$,
$\varphi(x,\phi)\sim\varphi_0(\phi)\sim-Q\phi$; at higher orders, $Q$ depends
on $\phi$).
$\beta^i$ are the {\it exact} beta functions in (\ref{laura}) of the theory
without gravity, that is, the graviton and tachyon beta functions of the sigma
model
without the additional target space coordinate $\phi$.
(\ref{sue}) describes the damped/antidamped motion of a particle in theory
space with the beta functions
as a driving force.
We refer to \cite{st} for details.\footnote{(\ref{sue}) has only been derived
for graviton- and $B$-field
perturbations there, but the extension to tachyon perturbations is
straightforward.}

Now, $\phi$ is related to renormalization group time $\tau$ by the
$x$-independent part of the
tachyon $T(x,\phi)=T(\phi)$, which is the dressed area operator
(see previous subsection and, e.g., \cite{das,polch,tseyt,ellis}).
For the $c=1$ model,
the relation between
$\phi$ and $\tau$ might be nonlinear (see \cite{polch}), but for the minimal
models ($c<1$), $T(\phi)$ has the simple form
$$T(\phi) \sim \mu\{ e^{\alpha\phi} + c^i_{\mu k}\times O(\lambda) +
O(\lambda^2)\},$$
where the second term corresponds to possible corrections of order $\mu\lambda$
that might arise if there are
nontrivial OPE coefficients $c^i_{\mu k}$ in the OPE of the cosmological
constant with one of the almost marginal
operators $V_k$. However, those coefficients are known to be zero for our
models. Therefore
\ba\tau \sim  -{\alpha\over2}\phi + O(\lambda^2).\la{barbara}\ea
With this identification, the string equations of motion become the
renormalization group flow equations.
If $\dot\lambda$ is of order $\lambda^2$ (as in (\ref{lay}) and as will be
assumed below),
the $\dot\lambda^2$ terms in (\ref{sue}) are of order $\lambda^4$ and can be
ignored at cubic order.
Likewise, the $O(\lambda^2)$ terms in $Q$
and the $O(\lambda^2)$ terms in (\ref{barbara}) can also be ignored
in the first equation of (\ref{sue}).
Thus, at least for $\kappa^2=0$ and $c<1$, gravity simply modifies the flow
equation (\ref{lay}) to
the second order differential equation
\ba {\alpha^2\over4}\ddot{\lambda^i}-{\alpha\over2}{Q}\dot{\lambda^i} =
\pi c^i_{jk}\lambda^j\lambda^k + \pi
d^i_{jkl}\lambda^j\lambda^k\lambda^l+...,\la{lisa}\ea
up to nonuniversal higher-order terms.\footnote{A possible small dimension
$\epsilon_i$ of $\lambda^i$ can be
absorbed in the $\lambda$ corresponding to the kinetic term for $x$.}
This is the precise form of the statement in point 1 of the introduction.
Now, one is interested in solutions of (\ref{lisa}) that also obey a
first-order flow equation
\ba \dot\lambda^i =
\pi {\tilde c}^i_{jk}\lambda^j\lambda^k +\pi {\tilde
d}^i_{jkl}\lambda^j\lambda^k\lambda^l+..., \la{mel} \ea
where one can call $\tilde c,\tilde d$ ``modified beta function coefficients''.
This ansatz guarantees that the flow without gravity is recovered in the
``classical limit''
of infinite negative central charge.
Differentiating (\ref{mel}), plugging it into (\ref{lisa}) and comparing the
quadratic and cubic
coefficients yields the ``gravitationally dressed'' coefficients:
\ba {\tilde c}^i_{jk} = -{2\over\alpha Q} c^i_{jk}, \ \ \ \ \ {\tilde
d}^i_{jkl}
 = -{2\over\alpha Q}( d^i_{jkl} - {2\pi\over
Q^2}{c}^i_{jm}{c}^m_{kl}).\la{sophie}\ea
The first result is the same as before. For the case of only one coupling
constant, where the coefficient $d$ is universal, the second result agrees
qualitatively
with a recent calculation of Dorn \cite{dorn}.\footnote{However, after fixing
normalizations
the results seem to differ by a factor 2. This might
signal a scheme dependence of $\tilde d$. I thank H. Dorn for pointing this out
to me (see ref. \cite{dorn2}) after this paper appeared.}
Since higher order beta function coefficients are not universal,
all information about the gravitational dressing of beta function
coefficients on a genus zero surface, at least in the vicinity of fixed points
with $c<1$, is thus encoded in the replacement
$$\dot\lambda^i \
\rightarrow\ {\alpha^2\over4}\ddot{\lambda^i}-{\alpha\over2}{Q}\dot{\lambda^i}\
$$
in the flow equation (\ref{laura}).

Next, one must ask whether this second order differential operator
is modified at genus-one level. In general, there are indeed terms of the form
\ba \kappa^2\dot\lambda,\ \ \kappa^2\ddot\lambda,\ \
\kappa^2\dot\lambda^2\la{jody}\ea
in the string equations of motion,
corresponding to terms like $\kappa^2R$ ($R$ is the target space curvature) in
the string effective action.
However, in the case discussed here, $\dot\lambda$ and $\ddot\lambda$ are at
most of order $\lambda^2$
or of order $\kappa^2$. Therefore the terms (\ref{jody}) are not relevant in
the present discussion, where at most cubic orders in the simultaneous
expansion in $\lambda$ and $\kappa$ are considered.
So in the following only the right-hand-side of (\ref{lisa}) will be modified
by topology fluctuations.

{}\subsection*{3. Node II and quantum mechanical flow}\scs{3}

In this section the scale dependence induced by node II is discussed to order
$\kappa^2$. Again, it must be remembered that
there are two scales: the fictitious background scale defined by $\sqhat$ and
the
physical scale defined by $\sqalpha$. Node II must be and will be seen to be
invariant under rescaling
of $\hat g$, but not under physical scale transformations, corresponding to
shifts of
the Liouville mode $\phi$.

Both the positively and the negatively dressed operators $V_i$ and $\bar V_i$
must be used in this section,
since they correspond to creation and annihilation operators in string theory
(see below).
The Liouville dressing of $\bar V_i$ grows faster
than $e^{-{Q\over2}\phi}$ in the infrared $\phi\rightarrow-\infty$. We believe
that this causes no problem at least as long as this operator
acts only on states with sufficiently positive Liouville momentum, so that it
never
creates states that ``do not exist'' in Liouville theory \cite{seipol}. Note
also
that negatively dressed operators can apparently be seen and studied in the
matrix models~\cite{bdks}.

 {}\subsubsection*{3.1. Massive modes}

The question is whether the integration over thin handles leads to new
divergences
that induce new dependence on physical scale transformations
$\phi\rightarrow\phi+{2\over\alpha}\tau$.
To study this, handles can be ``integrated out''. This leads to a bilocal
operator insertion on the
surface, whose behavior under constant shifts of $\phi$ can then be studied.

By ``integrating out handles'', the following is meant: surfaces are decomposed
into elementary vertices and
propagators (cylinders) as in string field theory \cite{zwie}, and cylinders
connecting a
surface with itself are replaced by
bilocal operator insertions. It must be emphasized that ``integrating out
handles'' does not mean reducing
higher-genus correlation functions to genus-zero correlation functions, since,
e.g., not all genus-1
surfaces can be regarded as a genus-0 surface plus a propagator. One still has
to integrate over genus-1
surfaces corresponding to the elementary genus-1 string field vertices. But
this will not lead
to any new logarithmic dependence on the world-sheet cutoff, so it will not
modify the flow of coupling constants.

To replace a cylinder by a bilocal operator insertion, one picks a complete set
of off--shell states propagating through the cylinder, consisting e.g. of
all scaling operators $\Phi_i(x)$ of the matter theory, properly normalized and
dressed with {\it arbitrary} Liouville momenta.
Then the cylinder ($\sim$ node II) can be replaced by the bilocal insertion of
{\it off-shell} operators
$$ \kappa^2\sum_i\int {d\epsilon\over2\pi}\ \  {1\over (L_0+\bar
L_0-2)_{i,\epsilon}}\ \ \int d^2z\ \Phi_i(z) e^{\epsilon\phi(z)}
\int d^2w\ \Phi_i(w) e^{(-Q-\epsilon)\phi(w)}$$
(such that a handle adds the amount $-Q$ to the Liouville momentum)
with inverse propagator
$$ (L_0+\bar L_0-2)_{i,\epsilon} = -(\epsilon+{Q\over2})^2+\omega_i^2.$$
The ``frequency'' $\omega_i$ has been defined in (\ref{rob}).
In a standard fashion, this sum over off-shell states propagating through the
cylinder can be replaced by a
sum over {\it on-shell} states by either using Feynman's tree theorem (see,
e.g., \cite{scherk}) or doing the
contour integral over macroscopic \cite{ssh} Liouville momenta $\epsilon$. This
yields in the case of
massive modes (``massive'' means $\omega_i>0$) an insertion
 \ba \sum_i {\kappa^2 \over 2\omega_i} \int d^2z\ V_i(z) \int d^2w\ \bar
V_i(w).\label{michelle}\ea
Since $V_i,\bar V_i$ are marginal,
(\ref{michelle}) is invariant under background scale transformations as it must
be,
except for the situation
where $V_i$ and $\bar V_i$ coincide; this will be taken care of in the next
section.
Under constant shifts $\phi\rightarrow\phi+{2\over\alpha}\tau$,
(\ref{michelle}) is multiplied by
$\exp\{(\alpha_i^++\alpha_i^-){2\over\alpha}\tau\}=
\exp\{-{2\over\alpha}Q\tau\}$. Thus,
all scale dependence that comes from massive states ($\omega_k>0$)
propagating through node II can be absorbed in a running topological coupling
constant
\ba \kappa^2 = \kappa^2(\tau) = \kappa^2_0\ \exp\{{2Q\over\alpha}\ \tau\}, \ \
\ \ \hbox{$\kappa^2_0 \equiv $ value at $\{\tau=0\}$}. \la{sam}\ea
Note that the ``dressed'' dimension $\vert{2Q\over\alpha}\vert$ of the string
coupling constant diverges in the weak gravity limit $c\rightarrow-\infty$, and
that $\kappa^2$ is dimensionless for $c=25$.

 {}\subsubsection*{3.2. Massless modes}

Let us now consider isolated ``massless'' modes ($\omega_k=0$) that propagate
through
node II. Massless modes correspond to dressed matter primary fields $\Phi_k$
with dimensions
\ba h_k\ \ =\ \ 2\ - \ {Q^2\over4}\ .\la{mia}\ea
Actually, in the $c\le1$ models there is only one example -- the cosmological
constant in
the $c=1$ model on a circle. $\Phi_k\propto1$ in this case. But the following
discussion will be kept general,
since it should also apply to more general models like supersymmetric theories
with $\hat c=9$ that may contain other isolated states with $\omega_k=0$.

Taking the limit $\omega_k\rightarrow0$ in (\ref{michelle}) yields a divergent
factor $\omega_k^{-1}$.
However, it must be remembered that $\omega_k$ is the Liouville momentum of the
state $|k>$
and the Liouville mode lives in a box $\phi_{ir}\le\phi\le\phi_{uv}$.
Here, $\phi_{ir}$ is the world-sheet infrared cutoff $-(\log\mu) /\alpha$
(whose value does not matter)
and $\phi_{uv} = (2\log a)/\alpha$ is the world-sheet ultraviolet cutoff.
Therefore $\omega_k$ cannot quite
become zero; instead, it should be replaced by its smallest possible value as
in \cite{kle}, i.e.
\ba{1\over\omega_k}\ \ \rightarrow\ \ \ \sim\ {|\phi_{ir}-\phi_{uv}|}\ \ =\ \
{1\over\alpha} (2\log a +\log\mu).\la{harry}\ea
The overall coefficient, which is not reliably fixed by this qualitative
argument, will be determined later.
So from taking the limit $\omega_k\rightarrow0$ in (\ref{michelle}) one learns
that,
under a rescaling $a\rightarrow ae^\tau$ of the physical cutoff,
isolated states with $\omega_k=0$ propagating through node II lead
to an operator insertion
\ba\sim\ \  \tau\ \ {\kappa^2\over\alpha} \int V_k \int V_k\ \ \ \ \
\hbox{with}\ \ \ \ \
V_k = \Phi_k e^{-{Q\over2}\phi}\ .\la{john}\ea
This bilocal scale dependence is independent of the value of $\mu$. It has its
origin in the fact
that $\phi$ is bounded by the physical cutoff $a$.
$\kappa^2$ depends on $\tau$ as in (\ref{sam}) to absorb the constant shift of
$\phi$ in $V_k$ that comes with the rescaling of the cutoff. Note that
there is no dependence on the fictitious cutoff $\hat a$, which
can been taken all the way to zero:
the integral over the node length $l$ decays exponentially though slowly (at
rate $\sim\vert\log a\vert^{-1}$)
and thus need not be cut off by $\log\hat a$.\footnote{While this paper was
being completed, an interesting
preprint
\cite{fisch} appeared in which - in a similar situation - bilocal divergences
containing the fictitious cutoff $(\log\hat a)$ and a
cancellation of the corresponding dependence on the fictitious scale are
discussed.}

The result (\ref{john}) is also plausible from a different (though related)
viewpoint. In the case $\omega_k=0$,
the two conjugate dressings of the operator $\Phi_k$ are $e^{-{Q\over2}\phi}$
and $\phi\ e^{-{Q\over2}\phi}.$
This suggests that integrating out node II produces a term proportional to
$${\kappa^2}\int \Phi_k\ e^{{Q\over2}\phi}\int \Phi_k\ \phi\ e^{{Q\over2}\phi}\
,$$
plus possibly a divergent term proportional to $\kappa^2\int\Phi_k
e^{{Q\over2}\phi}\int\Phi_k e^{{Q\over2}\phi}$ that transforms trivially under
constant shifts of $\phi$.
Shifting $\phi$ then yields a scale dependence of the form (\ref{john}).

Clearly, it is not possible to absorb the bilocal insertion (\ref{john}) in a
running coupling
constant $\lambda^k(\tau)$. But one can consider a Gaussian distribution with
width $\sigma$
in the space of theories parametrized by $\lambda^k$ (compare e.g. with
\cite{fried,cole,dass}). I.e., one can consider the ``averaged'' partition
function
$$ Z\ =\ \int d\lambda^k\ {1\over{\sqrt{2\pi}}\ \sigma}\exp\{-{1 \over
2\sigma^2}\lambda_k^2\}\ <\ e^{\lambda^k\int V_k} > .$$
The correlator on the right-hand side represents the partition function of the
original theory perturbed by $\lambda_k$.
It is assumed that $\sigma^2$ is of order $\kappa^2$. Performing the integral
over $\lambda^k$ yields
\ba Z\ =\  <\ \exp\{\ {\sigma^2\over2} \int V_k\int V_k\  \} > \  \
\sim\  \ <\ 1 \ +\  {\sigma^2\over2} \int V_k\int V_k\  +\ O(\kappa^4)\  >
.\la{barb}\ea
If one now introduces a ``running width'' with some initial value $\sigma_0$,
\ba\sigma^2\ \equiv\ \sigma^2(\tau)\ \sim\ (\ \sigma^2_0 -  \tau\
{2\over\alpha}\kappa_0^2\ )\ e^{{2Q\over\alpha}\tau},\la{barba}\ea
then (\ref{barb}) is independent of physical scale transformations;
in particular, the $\tau$-dependence of the second term in (\ref{barba})
cancels that of (\ref{john}).
Thus, the bilocal insertion (\ref{john}) is absorbed by letting the
distribution of theories spread under scale transformations in the direction of
isolated massless modes.\footnote{on top of the ordinary
$e^{{2Q\over\alpha}\tau}$ dependence of (sigma)${}^2$ due to the running
$\kappa^2$}
More precisely, it spreads towards the infrared, corresponding to decreasing
$\tau$.

This example illustrates that in the presence of isolated massless modes and
handles there is no ``classical'' renormalization group trajectory
$\lambda^k(\tau)$
 that describes
the same theory at different scales. Instead, averages, or ``wave packets'' of
theories must be considered
and the parameters that characterize their shape can also ``run''. At higher
orders in $\kappa,\lambda$ we expect new bi- and multilocal scale dependence,
signaling a more complicated ``running shape''.

That there is no classical renormalization group trajectory might in fact have
been guessed from the analogy between the flow on genus-zero random
surfaces and classical string theory that was discussed in the previous
section: since handles correspond
to string loops, the flow on higher genus surfaces should consequently be
described by quantum
string theory. The models discussed here have a discrete set of states;
therefore one arrives at quantum
mechanics rather than field theory. In more general models, the flow should be
described by an
effective quantum field theory for massless modes.

Having convinced ourselves that this ``quantization'' of the renormalization
group flow is indeed necessary, namely - to O($\kappa^2$) - in order
to absorb the scale dependence of the bilocal operator insertion (\ref{john}),
we can now fix the
proportionality constant in front of this insertion to be 1. This ensures that
the wave packet spreads at the rate
expected from quantum mechanics, $\sigma^2\sim\hbar t\sim\kappa^2\phi$.
$\sigma^2(\tau)$ in (\ref{barba}) then obeys the
flow equation
$$ {\alpha^2\over4}\ddot{\sigma^2}\ -\ {\alpha\over2}Q\dot{\sigma^2}\ \ =\ \
{Q}\ \kappa^2.\ $$
Comparing with (\ref{lisa}), one can regard $Q\kappa^2$ as the beta function
for the (width)${}^2$.

It should be emphasized
that in the presence of gravity the operators that produce bilocal logarithmic
divergences
are {\it not} the marginal operators of the matter theory with weight $h_k=2$,
as one might have expected,
but those with shifted weight (\ref{mia}).
In particular, because of this shift the radius $R$ in the $c=1$ model on a
circle coupled to gravity
is a superselection parameter: the wave packet does not spread out over $R$,
although the corresponding
operator $(\partial x)^2$ is marginal ($h=2$).
This ``failure of background independence'' has been noted in the matrix models
\cite{ssh}.

It is also noteworthy that the scale dependence of (\ref{john}) is linear in
$\tau$ and does not
behave like ${\sqrt\tau}$, as one might have expected since one momentum degree
of freedom
(the Liouville momentum) is integrated over.

{}\subsection*{4. Node III and loop-corrected beta functions}\scs{4}

We now turn to node III, which is well-known to modify the string equations of
motion
- this is the Fischler-Susskind mechanism \cite{fs} which has been amply
discussed in the literature
(see e.g., \cite{polski,cks,tsey,oog}).
Here we only discuss its implications for the flow in the $c=1$ model on a
circle
and point out some indirect matrix model evidence for this mechanism. In the
next section its
relation with the higher-genus string field vertices is pointed out.

 {}\subsubsection*{4.1. The Fischler-Susskind effect in the $c=1$ model}

Node III can be regarded as node I with the three-punctured sphere replaced
by a one-punctured torus (see the drawings in the introduction).
If the matter theory has a marginal operator $\Phi_i$, then in
analogy with (\ref{jimmy}) the corresponding state $|i>$ propagating through
node III yields the
logarithmically divergent local operator insertion
\ba-\ (\log \hat a)\ \pi\kappa^2\rho^i\int\bar V_i\  \la{isabelle}\ea
on the surface below, where the operators $V_i\sim\Phi_i,\bar
V_i\sim\Phi_ie^{-Q\phi}$ are normalized
to have unit two-point function on the sphere and the genus-one one-point
function
\ba \rho^i\ =\ {1\over v}<\int V_i>_{g=1} \ea
is the analog of the OPE coefficient $c^i_{jk}$ in the case of node I.
$v$ is the integral over possible zero modes of $\phi$ and the matter fields on
the torus.\footnote{The
zero mode integrals are overall integrals that must be divided out: the path
integral
factorizes into integrals over fields $x(\sigma_1)$ on the main surface
$\Sigma_1$, fields $x(\sigma_2)$ on the
surface $\Sigma_2$ that splits off and the zero-mode $x_0$:
${\cal D}x\ \rightarrow\ {\cal D}x(\sigma_1)\ {\cal D}x(\sigma_2)\ {d}x_0.$
Only the integral over $x(\sigma_2)$ is performed in replacing $\Sigma_2$ by an
operator insertion.}
Since the torus adds the amount $-Q$ to the Liouville background charge
the induced operator in (\ref{isabelle}) is the ``wrongly dressed'' one $\bar
V_i$.

In the $c\le1$ models, the only nonzero $\rho_i$ occurs for the (normalized)
operator $V \equiv{1\over2\pi\alpha'} (\partial x)^2$
in the $c=1$ model
(at higher genus, $\rho_i$ is also nonzero for the operator $R^{(2)}$; see
below).
If $x$ is compactified on a circle of radius $R$ one finds, setting $\alpha'=2$
as in section 2:
\ba\rho\
 =\ -{1\over48\pi} (1-{2\over R^2})\ .\la{astrid}\ea
Note that $\rho=0$ at the self-dual radius $R=\stwo$. To obtain (\ref{astrid}),
one observes that
\ba<{1\over2\pi\alpha'} \int(\partial x)^2>_{g=1}\ =\ -R{\partial\over\partial
R}\ Z_{g=1}\ .\la{gabi}\ea
This can be seen by redefining $x\rightarrow y=x/R$ in the
torus partition function (proportional to the negative free energy) \cite{kle}
$$ Z_{g=1}\ =\ \int {\cal D}x\ \exp\{-{1\over4\pi\alpha'}\int(\partial
x)^2+...\}
\ =\  {1\over12\stwo}\ ({R\over{\sqrt{\alpha'}}}+{{\sqrt{\alpha'}}\over R})\
\vert{\log\mu\over\alpha}\vert\ \ .$$
The volumes of the $\phi$ and $x$ zero modes in this model are given by
\ba v\ =\ \vert{{\log\mu\over\alpha}}\vert\ \times\  2\pi R.\la{assi}\ea

Let us for now ignore node II, which will be included in the next section.
(\ref{isabelle}) spoils the background scale invariance of the world-sheet
theory.
Similarly as in the case of node I, scale
invariance must be restored by adding a term of the form $\phi\bar V$
to the world-sheet action.
In the $c=1$ case this yields in analogy with (\ref{mary}) the kinetic term for
$x$ to order $\kappa^2$,
\ba \lambda_0\ {(\partial x)^2\over4\pi} \ -\ {\pi\over Q} \kappa^2 \rho\
{(\partial x)^2\over4\pi}\ \phi\ e^{-Q\phi}\ea
with $\lambda_0={1\over2}$.
The minus sign arises, because
$(L_0+\bar L_0-2)\ \phi\bar V_i = +2\omega_i\bar V_i$, as opposed to
(\ref{dora}).
This sigma model background solves the string equations of motion with
cosmological constant
\cite{fs}.\footnote{Note that this background does {\it not} describe a black
hole, despite of its
similarity to the black--hole operator $(\partial x)^2\ e^{-Q\phi}.$
Indeed, the ADM mass is zero \cite{guig}.}
Constant shifts
$\phi\rightarrow\phi+{2\over\alpha}\tau$ can now be absorbed in
\ba\kappa^2(\tau)\ =\ e^{{2Q\over\alpha}\tau}\ \kappa_0^2\ \ ,\ \ \ \
\lambda(\tau)\ =\ \lambda_0 +{2\over\alpha Q}{\pi}\rho\ \kappa^2(\tau)\ \tau\
,\la{silke}\ea
valid near $\lambda\sim{1\over2}$.
Note that the factor $2/(\alpha Q)$ differs from that in (\ref{layla}) only by
the minus sign.
There is no simple first order equation for $\lambda(\tau)$, but it
obeys a second order flow equation analogous to (\ref{lisa}) with
$c^i_{jk}\lambda^j\lambda^k$ replaced by $\rho^i\kappa^2$:
\ba {\alpha^2\over4}\ddot\lambda-{\alpha\over2}Q\dot\lambda\ =\ \pi\ \rho\
\kappa^2\ .\la{mali}\ea
This describes a damped motion towards the ultraviolet (remember that
$\alpha<0$), or equivalently
an antidamped motion towards the infrared.
Instead of keeping the radius $R$ fixed and letting $\lambda$ run,
it is more illuminating to absorb the change in $\lambda(\tau)$ in a
redefinition of $x$. Then
$R\rightarrow R{\sqrt{\lambda(\tau)/\lambda_0}}$
becomes $\tau$-dependent: using (\ref{silke}) and (\ref{astrid}), one finds the
flow equation
$$ {\alpha^2\over4}\ddot R-{\alpha\over2}Q\dot R\ =
\ -{1\over48}(R-{2\over R})\kappa^2\ +\ O(\kappa^4)\ .$$
This equation shows that there is a fixed point at the self-dual radius
$R=\stwo$.
Linearizing around it,
$R=\stwo+r,r\ll1$, one finds that this fixed point is unstable in the infrared
($\tau\rightarrow-\infty$):
the corresponding beta function (i.e., the right-hand side) is
$-{1\over24}\kappa^2 r$,
so $r$ is relevant (the fact that $\kappa^2=\kappa^2(\tau)$ does not change
this conclusion).
Here it is assumed that vortices are suppressed.
Vortices, if allowed, give additional contributions to the beta functions
\cite{gkl}; it is hard to compare
them with the $\rho\kappa^2$ term, since $\kappa^2$ grows exponentially.

 {}\subsubsection*{4.2. Comparison with the matrix model}

One might worry that - due to subtleties of Liouville theory - perhaps
Fischler-Susskind mechanisms are absent in the $c=1$ model coupled to gravity.
Is it possible to confirm their presence in the $c=1$ matrix model?
In the matrix model, the Fischler-Susskind effect should show up as a genus-1
effect, i.e. at first
order in the double scaling variable.
Therefore one should not expect to see the flow of the radius directly in the
genus expansion of the matrix models: instead of observing a
state $\vert(\partial x)^2\ e^{-Q\phi}>$ propagating through node III, one
should observe the state
$\vert(\partial x)^2>$ propagating in the opposite direction, corresponding to
node I.
Now, suppose two marginal external tachyon operators $\exp(\pm
i{2\over{\sqrt{\alpha'}}}x)$
(whose OPE contains the operator $(\partial x)^2$) are inserted into the sphere
as drawn below. This is just the
situation in which the propagation of the state $\vert(\partial x)^2>$ through
the node
leads to the Kosterlitz-Thouless transition on the torus!
It has indeed been observed in the matrix models that - not surprisingly - this
transition takes
place on surfaces of arbitrary genus \cite{gkl,moo}.

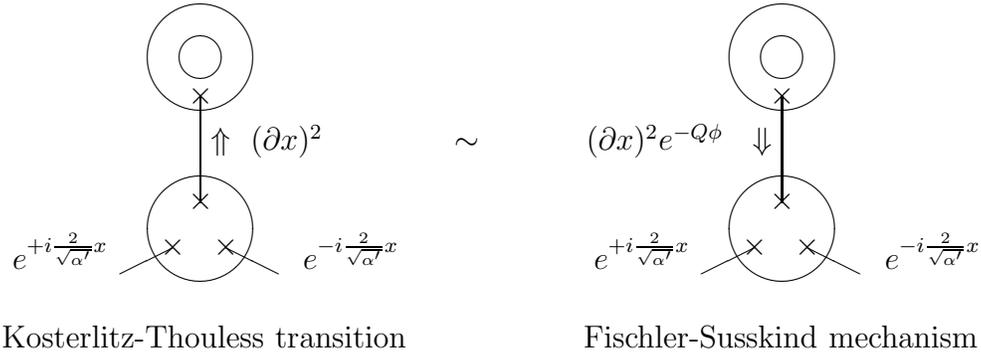
\begin{picture}(400,160)
\put(50,35){ \begin{picture}(100,100) \put(50,20){\circle{40}}
\put(50,85){\circle{40}} \put(50,30){\line(0,1){40}}
 \put(50,85){\circle{17}}
  \put(45.4,27.5){$\times$}  \put(45.4,67.5){$\times$}  \put(35,10){$\times$}
\put(55,10){$\times$}
  \put(39.6,13){\line(-2,-1){20}}  \put(59.6,13){\line(2,-1){20}}
\put(50,50){${\ \Uparrow}\  \ (\partial x)^2$}
\put(-25,5){
$e^{+i{2\over{\sqrt{\alpha'}}}x}$}
\put(85,5){
$e^{-i{2\over{\sqrt{\alpha'}}}x}$}
\put(-25,-25){Kosterlitz-Thouless transition}  \end{picture}}

\put(270,35){ \begin{picture}(100,100) \put(50,20){\circle{40}}
\put(50,85){\circle{40}} \put(50,30){\line(0,1){40}}
 \put(50,85){\circle{17}}
  \put(45.4,27.5){$\times$}  \put(45.4,67.5){$\times$}  \put(35,10){$\times$}
\put(55,10){$\times$}
  \put(39.6,13){\line(-2,-1){20}}  \put(59.6,13){\line(2,-1){20}}
\put(-24,50){$ (\partial x)^2e^{-Q\phi}\ \ \Downarrow$}
\put(-25,5){
$e^{+i{2\over{\sqrt{\alpha'}}}x}$}
\put(85,5){
$e^{-i{2\over{\sqrt{\alpha'}}}x}$}
\put(-25,-25){Fischler-Susskind mechanism}  \end{picture}}

\put(200,85){$\sim$}

\end{picture}

More generally, in the matrix model results Fischler-Susskind effects should be
manifest in the form of
poles in the higher-genus two-point function that are
due to the propagation of the on-shell states through node III.
If the torus is replaced by a surface of genus greater than one, contributions
to the Fischler-Susskind mechanism
come not only from the trace of the graviton $|(\partial x)^2>$ but also from
the zero-momentum
dilaton $|R^{(2)}>$ \cite{cks,polski,tsey}.
It is indeed possible to confirm the presence of poles due to these states (or,
equivalently,
their wrongly dressed counterparts) propagating through the node.\footnote{I
thank Igor Klebanov for pointing this out to me.}
To this end, consider the poles in the above genus-$g$ two-point function
of the (normalized) tachyon in the vicinity of the Kosterlitz-Thouless momentum
$p={2\over{\sqrt\alpha'}}$,
$$G^{(2)}(p  , -p)
\ \sim\ {1\over\epsilon}\ \sum_gt^{1-g}\ G_g\ \ \ \hbox{+\ finite}\ \ \ \ \
\hbox{with}\ \ \ p={2\over{\sqrt\alpha'}}+\epsilon\ .$$
Here $t=(2\beta\mu)^2$ with $\kappa^2=\pi^3\beta^2$ has been defined.
We now switch to $\alpha'=1$ to conform with the
notation of ref. \cite{kle},
from which we obtain the residues of the $1/\epsilon$ poles:
\footnote{These poles are present in the matrix model results, though
``hidden'' in the sense that
they are exactly cancelled
by poles at discrete tachyon momenta as described in \cite{gkn}.
This cancellation be understood
as a combination of a coupling constant redefinition that is singular at
discrete momenta and
of the terms (\ref{ted}).
It does not affect the conclusions of the present discussion.}
\ba
 G_1 &=& -{1\over12}(R-{1\over R})\vert\log\mu\vert
\\ G_2 &=& {1\over6!}\ (-21R-{10\over R}+{7\over R^3})
\\ G_3 &=& {1\over7!}\ (-155R-{147\over R}-{49\over R^3}+{31\over R^5}).
\ea
One now easily checks that \cite{igor}
\ba G_g\ \ =\ \ (\ -R {\partial\over\partial R}\ +\ \chi\ )\ Z_g\ ,\ \ \
\la{beate}\ea
where
\ba\chi\ =\ {1\over4\pi}\int d^2\xi\ {\sqrt g}R^{(2)}\ =\ 2-2g\ \ \
\la{marianne}\ea
is the Euler characteristic and $Z_g$ is the genus-$g$ partition function
($Z=\sum_gt^{1-g}Z_g$) \cite{kle}:
\ba Z_1 &=& {1\over12}(R+{1\over R})\vert\log\mu\vert
\\ Z_2 &=& {1\over6!}\ (7R+{10\over R}+{7\over R^3})
\\ Z_3 &=& {1\over7!}\ (31R+{49\over R}+{49\over R^3}+{31\over R^5}).
\ea
In (\ref{beate}), $-R{\partial\over\partial R} Z$ corresponds to
the genus-$g$ one-point function of the trace of the graviton (see
(\ref{gabi})),
while $\chi Z$ corresponds to the genus-$g$ one-point function of
the zero momentum dilaton (see (\ref{marianne})).
In this way the residues of the poles in the tachyon two-point function
can indeed be identified as the contributions
from these two states propagating through the node connecting the genus-$g$
surface with the
sphere containing the tachyons.
The fact that the $R$-dependence of the residues precisely works out confirms
indirectly that
Fischler-Susskind mechanisms work according to the theory in the $c=1$ model
coupled to gravity.

{}\subsection*{5. Relation with closed string field theory}\scs{5}

Let us now assemble pieces of the previous discussion to formally describe the
flow on random surfaces with handles in terms of quantum mechanics in theory
space,
with a potential that is corrected order by order in genus by the elementary
higher-genus vertices
of closed string field theory.\footnote{For other discussions of relations
between
between the flow and string field theory see \cite{fried,dass} and \cite{bru}.
Different conclusions about renormalization group flows, string theory and
other
issues can be found in \cite{ellis}.}

 {}\subsubsection*{5.1. Canonical formalism}

To order $\kappa^2\lambda^0$, it has already been seen in section 3 that
`quantization' of the flow
is necessary in the case of massless $\lambda^i$ in order to absorb the scale
dependence induced by node II.
At higher orders in $\lambda$ it will be useful to quantize {\it all} coupling
constants. By this, the following is meant.
Consider first the operator insertions (\ref{michelle}). As is well-known (see,
e.g., \cite{cole}),
such bilocal expressions can be reproduced
by turning the coupling constants into quantum operators.
More precisely, consider first the linearized flow equations. They can be
regarded as the
linearized string equations of motion for $\lambda^i(\phi)$ as discussed in
subsection 2.2
(here, a dot means derivative with respect to $\phi\sim-{2\over\alpha}\tau$):
\ba\ddot\lambda^k +Q\dot\lambda^k = \beta^k = (h_k-2)\lambda^k.\la{leroy}\ea
These equations can be transformed into the familiar (euclidean) harmonic
oscillator equations by redefining
$$\chi^k\equiv e^{{Q\over2}\phi}\lambda^k\ \ \ \rightarrow\ \ \ \ddot\chi^k
=(h_k-2+{Q^2\over4})\chi^k = \omega_k^2\chi^k.$$
The coupling constants
$\lambda^i$ (or $\chi^i$) can now be replaced by operators $\hat\lambda^i$ with
free mode expansion
$$\hat\lambda_k^{free}(\phi) \sim  \hat a_k^\dagger\ e^{\alpha_k^+\phi} +
\hat{a}_k\ e^{\alpha_k^-\phi}.$$
$\hat a_k^\dagger, \hat{a}_k$ are creation and annihilation operators with
commutation relations
$$ [\hat a_i,\hat a_j^\dagger]\sim{\kappa^2 \over 2\omega_i} \delta_{ij},\ \ \
\ \
[\hat a_i,\hat a_j] = [\hat a_i^\dagger,\hat a_j^\dagger] = 0.\ $$
${\hat a}^\dagger$ and $\hat a$ create and annihilate strings.
If correlation functions are sandwiched between in-- and out--vacua $|0>$ and
$<0|$,
defined by $\hat{a}_k\vert 0>\ =0$ and $<0\vert \hat a_k^\dagger =0$,
then the insertions (\ref{michelle}) are reproduced by
contractions of the $\hat a_k$'s and $\hat{a}_k^\dagger$'s in
\ba & & <\ <0|\ \exp\{\ \int\hat \lambda^{free}_i(\phi)\ \Phi_i\ \}\ \ V_1 ...
V_n\ |0>\ >\\
&\sim& \ \ <\ <0|\ \exp\{\hat a_i^\dagger\int V_i +\hat{a}_i\int \bar V_i \}\ \
V_1 ... V_n\ |0>\ >\ .\label{yvonne}\ea
The meaning of (\ref{yvonne}) can be made clear in the Schr\"odinger picture:
``Integrating out handles'' amounts to introducing random fluctuations
of the coupling constants around the fixed point, with a Gaussian distribution
of width
$\sigma\sim({\kappa/{\sqrt{\omega_i}}})$.
Unlike in the massless case $\omega_i\rightarrow0$, for $\omega_i>0$ this
distribution does not spread
under scale transformations,
apart from the overall spread that can be absorbed in the running topological
coupling constant $\kappa^2$.

Note that this  ``quantization'' procedure of the flow is also in accord with
the matrix model results
of \cite{berkle}. There it was shown that
indeed the torus free energy of a $(p,q)$ minimal model
can - after zeta function regularization and dividing out by the Liouville
volume - be written as the
sum over the (real time) harmonic oscillator ground state energies of all the
modes $\lambda^k$ with
the same $\omega_i$ that have been defined above:
$$F\ \sim\ \sum_i{\omega_i\over2}\ =\ - {(p-1)(q-1) \over 24(p+q-1)}.$$
In view of the interpretation of time $\phi$ as the scale factor, this could be
regarded as a
``zero-point central charge'' of conformal field theories in the presence of
handles.

Now, in the presence of interactions $c^i_{jk}$, the quadratic beta function
coefficients in (\ref{lisa}) must be included
in the equations of motion (\ref{leroy}). Solving them yields the
$O(\kappa^2\lambda)$
generalization of (\ref{yvonne}), dropping the operator insertions (compare
with (\ref{ted})):
\ba <\ <0|\exp\{ \hat a_i^\dagger\int V_i+\hat{a}_i\int \bar V_i +
  \pi{c^j_{ik} \over 2\omega_j}\hat a^{i\dagger} \hat a^{k\dagger} \int \phi
V_j
 -\pi {c^{\bar k}_{i{\bar j}} \over 2\omega_k}\hat a^j\hat a^{i\dagger}\int
\phi \bar V_k
\ +\ ...\ \} |0>\ >. \la{peter} \ea
Here it has been taken into account that there are nontrivial OPE coefficients
$$V_i(z)\bar V_j(0)\sim{1\over |z|^{2}}c^{\bar k}_{i\bar j} \bar V_k+...
\ \ \ \ \ \ \ \hbox{with}\ \ \ \ \ c^{\bar k}_{i\bar j} = c^{j}_{ik} \ .$$
It is interesting how (\ref{peter}) manages to be independent of the background
scale $\sqhat$.
Consider, e.g., the derivative of (\ref{peter}) with respect to $\hat
a^{i\dagger}$,
 \ba <\ \int V_i\ +\ \pi{\kappa^2\over2\omega_k}{c^j_{ik} \over 2\omega_j} \int
\phi V_j\int \bar V_k
 \ -\ (j\leftrightarrow k)\ \ \ + \ \ \hbox{other contractions}\ >.\la{matt}
\ea
The second and third terms come from the contraction of linear and quadratic
terms in the exponential.
Now, $V_i$ in (\ref{matt}) may collide with one of the operators of the bilocal
insertions (\ref{michelle})
(which are produced by contractions of the linear terms in (\ref{peter})):

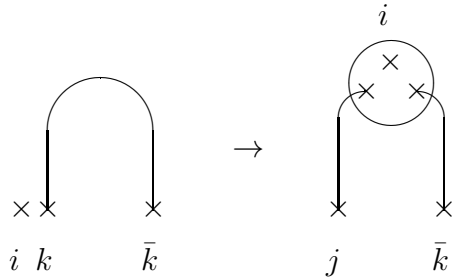
\begin{picture}(400,120)
\put(100,0){ \begin{picture}(100,100)
\put(50,30){\oval(40,100)[t]}
  \put(15.4,27.5){$\times$}
  \put(100,50){$\rightarrow$}
  \put(15.4,7.5){$i$}  \put(25.4,7.5){$k$}  \put(65.4,7.5){$\bar k$}
  \put(25.4,27.5){$\times$}  \put(65.4,27.5){$\times$}    \end{picture} }

\put(210,0){ \begin{picture}(100,100)
  \put(40,30){\oval(20,90)[tl]}
  \put(60,30){\oval(20,90)[tr]}
  \put(50,78){\circle{30}}\put(36,72){$\times$}
  \put(55,72){$\times$}\put(45,83){$\times$}
  \put(45,100){$i$} \put(25.4,7.5){$j$}  \put(65.4,7.5){$\bar k$}
  \put(25.4,27.5){$\times$}  \put(65.4,27.5){$\times$}    \end{picture} }
\end{picture}

\noindent
This situation leads to the standard pole in string amplitudes that is due to
the ``pair production'' of ``particle'' $j$ and ``anti--particle'' $\bar k$. On
the world-sheet,
this pole shows up as a logarithmically divergent bilocal operator insertion
that depends on the
fictitious scale through the cutoff $\hat a$:
 \ba \ \pi\ (\log \hat a)\ {\kappa^2\over
2\omega_k}{c^j_{ik}\over2\omega_j}\int V_j\int \bar V_k
 \ \ \ \ \ \ +\ \ \ \ \ (j\leftrightarrow k).\la{mack}\ea
It can be found by either integrating over the moduli or by using the OPE of
$V_k$ with $V_i$.
Using (\ref{dora}) (and its counterpart for $\bar V$), one sees that the scale
dependence of (\ref{mack})
is now precisely cancelled by that of the second and third terms in
(\ref{matt}).

 {}\subsubsection*{5.2. Closed string field vertices}

We can now show
how the higher-genus string field vertices formally enter
the flow equations. The point is that
part of node III corresponds to the situation where the two ends of node II
coincide with
each other, forming a three-vertex with two legs connected by a propagator (see
figure below).
The corresponding scale dependence is {\it already} cancelled if the
$\lambda^i$ are turned into operators in order to account for node II.
Namely, in (\ref{peter}) the contraction of the last term with itself
corresponds to inserting a propagator on the 3-vertex. This
yields already at order $\kappa^2\lambda^0$ an insertion
$$ \ - \sum_{i,j,k} {\pi\over2\omega_k}c^{\bar k}_{i\bar j}\ <0|\hat a_j\hat
a_i^\dagger|0>\
<\int\phi\bar V_k>\ \ \sim \ \ -\sum_{i,k} {\pi\kappa^2\over
4\omega_i\omega_k}c^i_{ik}\ <\int\phi\bar V_k>\ .$$
Since $(L_0-\bar L_0-2)\phi\bar V_k=-2\omega_k V_k$, this
cancels part of the scale dependence (\ref{isabelle}) that is due to node III.
Only the scale dependence from the remainder of node III,
$$(\nu_{1,1})_i \ \sim\ \rho_i + \sum_l{1\over2\omega_l}c^l_{il}, $$
must then be cancelled by adding an $O(\kappa^2)$ term to the beta functions as
in (\ref{mali}).
$\nu_{1,1}$ is the elementary genus-1 1-vertex of closed string field theory
\cite{zwie} -
the part of the moduli space of a one-punctured torus that cannot be obtained
by
connecting a genus-zero vertex with a propagator.

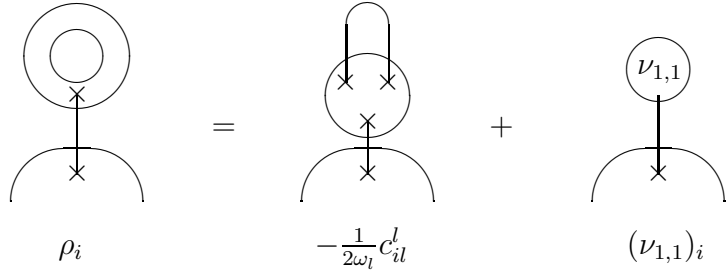
\begin{picture}(400,125)

\put(40,10)
 { \begin{picture}(100,100) \put(50,20){\oval(50,40)[t]}
\put(50,75){\circle{40}} \put(50,75){\circle{18}}
  \put(50,30){\line(0,1){30}} \put(45.4,27.5){$\times$}
\put(45.4,57.5){$\times$}  \put(43,-0){$\rho_i$}  \end{picture} }

\put(150,10)
 { \begin{picture}(100,100)
  \put(50,20){\oval(50,40)[t]}
  \put(50,65){\oval(16,60)[t]}
  \put(50,60){\circle{30}} \put(50,30){\line(0,1){20}}
  \put(45.4,27.5){$\times$}  \put(45.4,47.5){$\times$}  \put(37,62){$\times$}
\put(53,62){$\times$}  \put(30,-0){$-{1\over2\omega_l}c^l_{il}$}
\end{picture}}

\put(260,10)
 { \begin{picture}(100,100) \put(50,20){\oval(50,40)[t]}
\put(50,70){\circle{25}}
  \put(50,30){\line(0,1){30}} \put(45.4,27.5){$\times$}
\put(42,68){$\nu_{1,1}$} \put(38,-0){$(\nu_{1,1})_i$ }  \end{picture} }

\put(145,55){=} \put(250,55){+}
\end{picture}

\noindent
The flow equations (\ref{lisa}) finally become, after rescaling the vertices
(for
almost marginal operators with dimensions $h_i\sim2$):
\ba {\alpha^2\over4}\ddot{\hat\lambda^i} - {\alpha\over2}Q\dot{\hat\lambda^i} =
(h_i-2)\hat\lambda^i + \pi c^i_{jk}\hat\lambda^j\hat\lambda^k+ \pi d^i_{jkl}
\hat\lambda^j\hat\lambda^k\hat\lambda^l+\kappa^2(\nu_{1,1})^i+...\la{sally}\ea
In fact, $c^i_{jk}$ and $d^i_{jkl}$ can be thought of as the genus-0 string
field vertices $\nu_{0,3}$ and $\nu_{0,4}$.
Those are the genus-0 3-point function and the part of the genus-0 4-point
function
that cannot be built by connecting two 3-point functions with a propagator (the
other part
is already taken care of by solving (\ref{sally}) for $\lambda^k$ to order
$\lambda^2$
and then plugging the result back into the quadratic term).
In this sense the flow is described, at least to this order, by canonically
quantized closed string field theory.

More generally, turning the coupling constants $\lambda^i$ into operators
should absorb all
scale dependence that arises from handles that are added to tree diagrams
(nodes II). The remaining scale dependence
must be cancelled by hand with the help of the elementary genus-$g$
$N$-vertices $\nu_{g,N}$.
At higher orders in $\lambda,\kappa$ those may also modify the time derivative
operator in (\ref{sally}),
as noted at the end of section 2.
Of course, a complete treatment of the BRST cohomology should involve the full
BV formalism as in \cite{zwie}.

{}\subsection*{6. Speculation and summary}

 {}\subsubsection*{6.1. Speculation for $\hat c\ge9$}

In the previous sections, the vicinity of fixed points with central charge
$c\le1$ has been discussed.
Let us now suggest how our discussion should be modified for models with
$c\ge25$ -- or rather, for
supersymmetric theories with $\hat c\ge9$ coupled to supergravity, with the
tachyon projected out.\footnote{For a discussion of the gravitational dressing
of one-loop beta functions in $N=1$ supersymmetric theories with $\hat c\le1$
(and of the absence of such a dressing in $N=2$ theories) see \cite{gri}.} The
fixed points with $\hat c\ge9$ can, e.g., be sigma models with Euclidean
signature; the conformal factor then becomes a timelike target space
coordinate.
\begin{enumerate}
\item
Concerning node II, since
these theories contain full-fledged target space fields, the flow should be
described by quantum field theory
rather than quantum mechanics.
Concerning node III, the genus-one tadpole effects should cancel due to
supersymmetry.
\item
In analogy with \cite{st} for the case $c\ge25$, we expect that for $\hat
c\ge9$
the flow to the infrared is damped rather than anti-damped. This is a result of
the fact that time is minkowskian for $\hat c\ge9$.
As a consequence, {\it both} Liouville-dressings {\it con}verge at an infrared
fixed point and
both correspond to equally good world-sheet operators, at least in the infrared
region.
Presumably this means that the general renormalization group trajectory is
a general string solution, which contains
both dressings. In the $c\le1$ models by contrast, the ``wrong'' Liouville
dressing which
corresponds to the divergent classical solution is suppressed in the path
integral (as is usual
for euclidean quantum mechanics).
In theories on a fixed lattice, of course, the coupling constants at one scale
must be uniquely
determined by the coupling constants at a different scale. But in our case the
scale is
a dynamical variable,
and it is no longer clear that a {\it unique} renormalization group trajectory
must pass through a given
point in theory space.

\item
Before settling down at an infrared fixed point, renormalization group
trajectories may oscillate around it.
If the central charge of the infrared fixed point is $\hat c=9$, we expect in
analogy with an example in \cite{st}
that the oscillations are no longer exponentially damped but decay much slower
in time, $\sim{1\over t}$.
They correspond to wave-like excitations of the target space fields.

\item
Furthermore, we expect that there are
two disconnected sectors of string solutions or renormalization group
trajectories - those corresponding to
euclidean and minkowskian target space signature. Static trajectories
corresponding to fixed points
 with $\hat c>9$ are in the
latter sector, while those corresponding to fixed points with $\hat c<9$ are in
the former sector.\footnote{Such static trajectories correspond to the
cosmological solutions of \cite{ant}.}
(fixed points with $\hat c=9$ can belong to both sectors, depending on how they
are approached).
If a theory starts from an ultravioulet fixed point with $\hat c\ge9$, it
cannot converge towards an
infrared fixed point with $\hat c<9$, as in \cite{st}.
One may speculate that the quantum analog of this statement is that string
vacua with $\hat c=9$ cannot decay.

\item
What are the infrared stable fixed points with $\hat c\ge9$? Apparently they
must obey
two conditions. First there must be no relevant operators
in the matter theory, or equivalently no tachyons in the string theory;
otherwise the fixed point is
not infrared stable. Second, it seems that the matter theory must be modular
invariant because
the moduli are integrated over - i.e., any theory coupled to gravity should at
least be equivalent to
a modular invariant theory coupled to gravity.
Combining both conditions, one concludes that the infrared fixed points
correspond to consistent string
theories!

\end{enumerate}

In view of this, it is of course very tempting to regard closed string field
theory
as the theory of the flow in the most general unitary theory that lives on a
(super)
random surface. One might speculate that this is a continuum limit
of some unknown statistical mechanical system - perhaps a $\hat c\ge9$
analog of the ensemble of large Feynman graphs of the matrix models.
Like any such system, it would flow to an infrared fixed point - a string
vacuum.
Since
renormalization group time is identified with real time in target space, this
flow
would have the interpretation
of a cosmological evolution.

In the vicinity of fixed points, it would make sense to describe the system as
a target space field theory.
If the infrared fixed point had $\hat c=9$, perhaps after tunneling down from
$\hat c\ge9$,
there would be oscillations of the flow around it.
Their spectrum could be compared with
the observed spectrum of elementary particles.
It must be pointed out, though, that if one took this seriously, one would have
to assume that the underlying statistical mechanical system is
huge: since time corresponds to the world-sheet scale, we would
presently be observing this system at scales of the order of at least
$$10^{10^{61}}\ $$
times its cutoff-scale, where $10^{61}$ is the age of the universe in Planck
units,
and in fact this number would be growing fast as ``scale'' goes by.

 {}\subsubsection*{6.2. Summary}

Let us summarize the effects of gravity on the renormalization group flow that
have been discussed here.
It has been seen that, due to fluctuations of the conformal factor,
at least for $c<1$ and up to cubic order in coupling constants,
the time derivative in the standard flow equation (\ref{laura}) is simply
replaced
by the second-order derivative operator
$$\dot\lambda^i\ \ \rightarrow\ \
{\alpha^2\over4}\ddot\lambda^i-{\alpha\over2}Q\dot\lambda^i\ .$$
To this order, this turns the flow equations into the string equations of
motion.
The following variations of this general theme of second-order flow equations
have been discussed (a dot means derivative
with respect to $-{\alpha\over2}\phi$):

First, for (almost) marginal operators the flow equations become
$$ {\alpha^2\over4}\ddot\lambda^i\ -\ {\alpha\over2}Q\dot\lambda^i\ =\
c^i_{jk}\lambda^j\lambda^k\ +\ d^i_{jkl}\lambda^i\lambda^j\lambda^k\ +\ ...$$
Picking the solution that also obeys a standard first-order equation yields the
gravitational dressing
(\ref{sophie}) of the
beta function coefficients $c^i_{jk}$ and $d^i_{jkl}$ for almost marginal
operators.
The result for the dressing of $c^i_{jk}$ agrees with the light-cone gauge
result. It would be interesting to also compute
the dressing of $d^i_{jkl}$ in light-cone gauge.

Second, in the presence of handles the
topological coupling constant $\kappa^2$ runs as in (\ref{sam}).
To lowest order in $\kappa^2$ it obeys the flow equation
$$ {\alpha^2\over4}\ddot{\kappa^2}\ -\ {\alpha\over2}Q\dot{\kappa^2}\ =\ 0 .$$
Furthermore, a curious phenomenon takes place if the matter theory has isolated
``massless'' states with dimension
$h_i = 2-{Q^2\over4}.$
Due to pinched handles, the `classical' renormalization group trajectory
$\lambda^i(\tau)$ is then replaced
by a ``wave packet'' of theories that spreads under scale transformations
in the directions corresponding to these $\lambda^i$.
Its width square $\sigma^2$ obeys the flow equation
$${\alpha^2\over4}\ddot{\sigma^2}\ -\ {\alpha\over2}Q\dot{\sigma^2}\ =\
Q\kappa^2 $$
with `beta function' $Q\kappa^2$.
More generally, in the presence of handles distributions
of theories must be considered instead of points in theory space. The moments
that describe their shape also become running coupling constants with flow
equations analogous to those for the $\lambda$'s.
Effectively, the flow is then described by a quantum theory for massless modes.
One wonders whether there are related phenomena in solid state physics or
statistical mechanics.

Third, beta functions are modified by Fischler-Susskind effects.
E.g., in the $c=1$ model on a circle
the radius becomes a running coupling constant, obeying the equation
($\alpha'=2$):
$$ {\alpha^2\over4}\ddot R\ -\ {\alpha\over2}Q\dot R\ =
-{1\over48}\kappa^2(R-{2\over R}) .$$
The self-dual radius corresponds to an ultraviolet stable fixed point.
The poles associated with the Fischler-Susskind mechanism that are due to the
trace of the graviton and the zero-momentum
dilaton propagating through node III precisely agree with the matrix model
results
at least up to genus 3.
After ``quantizing'' the flow to account for node II, i.e., after replacing the
coupling constants by operators, the loop corrections to the beta functions
reduce
to the elementary vertices $\nu_{g,N}$ of closed string field theory:
$$ {\alpha^2\over4}\ddot{\hat\lambda^i}\ -\ {\alpha\over2}Q\dot{\hat\lambda^i}\
\sim\ \beta^i\
+\ \kappa^2(\nu_{1,1})^i\ +\ \kappa^2(\nu_{1,2})^i_j\hat\lambda^j\ +\  ...$$
(only the case $\nu_{1,1}$ has been discussed explicitly. At higher orders, the
left-hand side might also be
corrected). This is in accord with the assertion
that the renormalization group flow in theories that live on random surfaces
with handles
is described by closed string field theory.
One might have thought that the
theory of the flow ins only ``half of string theory'', since the solutions with
negative Liouville dressing are forbidden; but if the argument in point 2 of
the previous subsection is correct, both dressings
should be allowed in the regime $\hat c\ge9$, where the flow to the infrared is
damped rather than anti-damped.

A very interesting question is whether the flow in the presence of handles
can also tunnel between different infrared fixed points (as has been speculated
in the last subsection). One could, e.g., imagine that
such tunneling induces a transition from the $c=1$ phase to the $c=0$ phase of
the O(2) model on a random surface.
Perhaps this can be checked in the corresponding matrix model \cite{gkl}.
It would be a nonperturbative effect of order exp(${-{1/\kappa^2}}$).
Another challenge of course is to understand the nonperturbative features of
order $\exp(-{1/\kappa})$ of the flow between
minimal models that have been observed in the matrix models \cite{inst}.

Finally, let us note that one can write down
fixed point conditions for the flow on random surfaces of arbitrary genus
by cancelling the scale-dependence from the various components of the boundary
of moduli space.
They are (somewhat) reminiscent of
the Virasoro constraints \cite{ver}, the master equation for $2d$ string theory
\cite{erik},
and the
holomorphic anomaly equation of \cite{vaf}. It is not clear to the author
whether there is any connection.

\vskip 10mm
\subsection*{Acknowledgements}

I would like to thank Sasha Polyakov for helpful discussions,
Igor Klebanov for useful suggestions and
the Princeton University Physics group for its encouraging interest in these
ideas during
a talk in February. This work is supported in part by ``Deutsche
Forschungsgemeinschaft''
and also by NSF grant PHY-9157482 and James S. McDonnell Foundation grant No.
91-48.

{}

\end{document}